\documentclass[prd,aps,eqsecnum,floatfix,nofootinbib,preprint,tightenlines]{revtex4}

\usepackage{latexsym}
\usepackage{graphicx}
\usepackage{multirow}
\usepackage[dvipsnames]{xcolor}

\def\bibi{\bibitem}

\def\a{\alpha}
\def\b{\beta}

\def\d{\delta}
\def\e{\epsilon}                % Also, \varepsilon
\def\g{\gamma}

\def\m{\mu}

\def\o{\omega}
\def\p{\pi}                     % Also, \varpi
\def\t{\tau}

\def\D{\Delta}

\def\G{\Gamma}

\def\P{\Pi}

\def\cbo{{\,\raise-.15ex\Sc [\,}}                       % curly "
\def\ddt#1{{\buildrel {\hbox{\LARGE .\kern-2pt.}} \over {#1}}}% double dot-over
\def\ie{\mbox{\it i.e.}}
\def\eg{\mbox{\it e.g.}}

\def\half{{1\over 2}}

\def\floatcaption#1#2{ \caption{ #2 \ [#1] \label{#1}} }
\def\floatcaption#1#2{ \caption{#2 \label{#1}} }

\def\bibi{\bibitem}    %% uncomment to suppresses citation labels

\def\ttl#1{{\it #1}}
\begin{document}

\begin{boldmath}
\begin{center}
{\large{\bf On the difference between Fixed-Order and Contour-Improved Perturbation Theory
}}\\[8mm]
Maarten Golterman,$^a$ Kim Maltman,$^{b,c}$ Santiago Peris,$^{a,d}$\\[8 mm]
$^a$Department of Physics and Astronomy, San Francisco State University,\\
San Francisco, CA 94132, USA\\
[5mm]
$^b$Department of Mathematics and Statistics,
York University\\  Toronto, ON Canada M3J~1P3
\\[5mm]
$^c$CSSM, University of Adelaide, Adelaide, SA~5005 Australia
\\[5mm]
$^d$Department of Physics and IFAE-BIST, Universitat Aut\`onoma de Barcelona\\
E-08193 Bellaterra, Barcelona, Spain
\\[10mm]
\end{center}
\end{boldmath}

\begin{quotation}

Using standard mathematical methods for asymptotic series and the large-$\b_0$ approximation, we define a Minimum Distance between the Fixed-Order perturbative series  and the Contour-Improved perturbative series in the strong coupling $\a_s$ for finite-energy sum rules as applied to hadronic $\t$ decays.
This distance is similar, but not identical, to the Asymptotic Separation of Hoang and Regner, which is defined in terms of the difference of the two series after Borel resummation. Our results confirm a nonzero nonperturbative  result in $\a_s$ for this Minimum Distance as a measure of the intrinsic difference between the two series, as well as a conflict with the Operator Product Expansion for Contour-Improved Perturbation Theory.

\end{quotation}

\section{\label{Intro} Introduction and motivation}

The determination of $\a_s$ from an analysis of the hadronic $\t$ decay has a very long and successful history, which is  linked to the use  of Finite Energy Sum Rules (FESRs) \cite{shankar}-\cite{BNP}. These FESRs rely on the analyticity of the  Vector (and Axial) two-point correlator to relate integrals of the spectral function to complex contour integrals of perturbation theory supplemented by the nonperturbative contribution from the Operator Product Expansion (OPE). Because the mass of the $\t$ lepton is not much larger than the scale of QCD, one of the problems that one has to face is that of the (non)convergence of the perturbative series, a property associated with renormalons \cite{Beneke:1998ui}. Even the OPE as a whole has been recognized as
constituting a nonconvergent expansion, and this property is at the origin of
(quark-hadron) Duality Violations \cite{PQW}-\cite{BCGMP}.

Although the QCD perturbative series is expected to be only asymptotic, multiple ideas have been explored to compensate for this lack of convergence by a reorganization of the terms in the series in order to help approach the final result at the  fastest possible  rate \cite{Caprini}-\cite{Abbas}. Two of these organizations of the perturbative series have been predominant in analyses of the hadronic $\t$ decay: Fixed-Order Perturbation Theory (FOPT) and Contour-Improved Perturbation Theory (CIPT) \cite{CIPT}-\cite{CIPT2}.

The essential difference between FOPT and CIPT is in the choice of the renormalization scale $\mu^2$ which is present in any perturbative calculation. Usually the natural choice for a QCD correlator with an external scale $s_0$ has been to take $\mu^2$ equal to this external scale to minimize the contribution of the ubiquitous logarithms of the form $\log( s_0/\m^2)$ appearing in the perturbative series. This is the choice made in FOPT. However,  the existence of the contour of integration in the complex plane in the FESRs also allows the possibility to make the choice $\mu^2=z$  to kill the logarithmic contribution of the form $\log(z/\m^2)$ along the whole contour of integration in the $z$ complex plane, and this latter choice is  the one made in CIPT.

The proponents of CIPT defend their position by arguing that this choice resums the perturbative series along the whole complex contour, resulting in a series which, at least for the lowest orders that have been calculated  in $\t$ decay so far, seems to be better behaved, \ie, with smaller term-by-term contributions. However, apart from its unique application,\footnote{To the best of our knowledge, a  choice like  CIPT has never been made in any  QCD analysis apart from its use in FESR
analyses. It is far from obvious whether the CIPT $\a_s$ is the same coupling as that extracted from an observable for which there exists no contour to consider.} the partial resummation carried out by CIPT is potentially suspect. As mentioned above, the QCD perturbative series is asymptotic and it turns out that  this partial resummation disrupts a cancellation between the leading perturbative orders from the Adler function and the orders originating from the contour of  integration. In fact, in the large-$\b_0$ approximation this cancellation is exact \cite{BJ}.

Recently, the seminal work of Ref.~\cite{Hoang:2020mkw} has substantially contributed to a clarification of the difference between FOPT and CIPT. These authors have pointed out that the Borel transform which generates the CIPT series is  different from that of the FOPT series, resulting in a irreducible difference, of $\mathcal{O}(e^{-c/\a_s})$ (with $\a_s$ the strong coupling and $c$ some positive constant), which they refer to as the ``Asymptotic Separation" (AS). Since the renormalon structure of the FOPT series, and in particular its associated ambiguities, match the contributions from the OPE as expected,\footnote{Any ambiguity generated by the perturbative series is expected to be cured by a contribution from the OPE, since the combined result should be unambiguous.} this extra nonperturbative contribution present in the CIPT series is in conflict with the OPE or, at least, with its standard version \cite{Shifman:1978bx},  in that it suggests the existence of nonperturbative terms which go beyond the OPE.

However, in spite of this, in some analyses it has been standard practice to assume the same OPE for FOPT and for CIPT, taking an average of the results obtained for $\a_s$ in both cases, and taking the difference as some sort of systematic error \cite{Pich:2022tca}. In view of the results of Ref.~\cite{Hoang:2020mkw}, this procedure now has turned out to be inconsistent.

What makes sense, once the source for this irregular behavior of the CIPT series has been identified, is to design a cure for it. Reference \cite{Benitez-Rathgeb:2022yqb} has presented a scheme to bring CIPT into agreement with the OPE, and the results obtained largely confirm the results obtained with the FOPT method.

The purpose of the present note is to discuss the difference between Fixed-Order Perturbation Theory (FOPT) and Contour-Improved Perturbation Theory (CIPT), following the analysis of Hoang and Regner \cite{Hoang:2020mkw}. However, in contrast to the approach of Ref.~\cite{Hoang:2020mkw}, which considers this difference at the level of a Borel transform of the contour integration generally used in $\t$-decay analyses, in this work we will consider the asymptotic series which results after the integration over this contour is carried out, order by order, at the level of the perturbative series expansion.  The result of this analysis confirms that there is a nonperturbative difference between the FOPT and CIPT asymptotic series, as pointed out in Ref.~\cite{Hoang:2020mkw}, and we agree with the main message of that reference, namely, that this nonperturbative difference is at odds with the OPE. In addition, we show how to reproduce the results of Ref.~\cite{Hoang:2020mkw} in a different way, at least in the large-$\b_0$ approximation. However, our new result for the difference at the level of the series is not the same as the AS obtained at the level of Borel sums in Ref.~\cite{Hoang:2020mkw}

In this work we will limit ourselves to the large-$\b_0$ case. We do this because this case is
considerably simpler while at the
same time containing the essential
ingredients, allowing for a clearer
discussion of the main results.   Therefore, let us define the variable $\a$ as
\begin{equation}
\label{alpha}
\a=\half\,\b_0\a_s\ ,
\end{equation}
with (using $\m\, d\a_s/d\m=-\b_0\a_s^2$)
\begin{equation}
\label{beta0}
\b_0=\frac{1}{\p}\left(\frac{11N_c}{6}-\frac{N_f}{3}\right)\ .
\end{equation}
The Renormalization Group equation in terms of $\a$ simply reads
\begin{equation}
\label{deralpha}
-z\frac{d}{dz}\a(z)=\a^2(z)\ .
\end{equation}
For future reference, let us note that  $N_f=3$ and $\a_s=0.3$ (the approximate value of $\a_s$ at the $\t$ mass) correspond to $\a\simeq 0.2$, which is the relevant value in the analysis of the hadronic decay of the $\t$ lepton. This is why in some examples we will use $\a=1/5$.

\begin{boldmath}
\section{\label{FOPTAdler} FOPT for the Adler function and arbitrary moment $x^m$}
\end{boldmath}
The Adler function, $D(z)$ posseses a renormalon structure \cite{Beneke:1998ui} which admits a Borel representation
given  by
\begin{eqnarray}
\label{Adler}
D(z)=-z\frac{d\P(z)}{dz}&=&\int_0^\infty d\o\, B(\o) \,e^{-\frac{\o}{\a(-z)}}\\
&=&\int_0^\infty d\o\, B(\o) \left(\frac{-z}{\m^2}\right)^{-\o} \,e^{-\frac{\o}{\a(\m^2)}}\nonumber
\end{eqnarray}
in terms of the Borel function, $B(\o)$, whose singularities closest to the origin are located at integer values of $\o$.   The second equation in Eq.~(\ref{Adler}) follows from Eq.~(\ref{deralpha}).
In Eq.~(\ref{Adler}), $\Pi(z)$ stands for the scalar function associated with the vacuum polarization tensor of the two-point vector correlator.

One distinguishes the ultraviolet (UV) renormalons, as those sigularities located at negative values of $\o$, and the infrared (IR) renormalons, with singularities located at positive values of $\o$.  In most of this work we will concentrate on the IR renormalons, as they will turn out to be mainly responsible for the FOP-CIPT difference, but we will also comment on the UV renormalons and clarify their role, if any,  in regard to this difference.

In general, these singularities are branch points but, in the large-$\b_0$ approximation, the IR renormalons are simply poles in the Borel function $B(\o)$: a simple pole at $\o=2$ plus an infinite sequence of double poles at $\o=3,4,5,...$. Therefore, we may concentrate on the contribution of an individual pole to the Adler function by considering the integral
\begin{equation}
\label{Adler2}
D(z)|_{p}=\int_0^\infty d\o\,\left(\frac{-z}{\m^2}\right)^{-\o}\,\frac{1}{(p-\o)^\g}\,e^{-\frac{\o}{\a(\m^2)}}\ .
\end{equation}
where, as we have discussed, $\g=1$ or $2$ in the large-$\b_0$ approximation.

At this level the distinction between FOPT and CIPT lies in the choice for the scale $\m$. In FOPT, this choice amounts to taking $\mu^2=s_0$, where $s_0$ is a typical external Euclidean scale in the problem which, in the particular case of the hadronic $\t$ decay, may be simply identified with $m_\t^2$. Because the $\t$ mass is not much larger than the hadronic scale, analyses of the decay of the $\t$ lepton traditionally use the so-called Finite Energy Sum Rules (FESRs), which consist of moments, $A_m(s_0)$, of contour integrals of radius $|z|=s_0$ around the origin of the complex plane. Therefore, we will study integrals of the type
\begin{equation}
\label{Amomdef}
A_m(s_0)=-\frac{1}{2\p i}\oint_{|x|=1}\frac{dx}{x}\,x^m\,D(xs_0)\ .
\end{equation}
Upon inserting (\ref{Adler2}) into (\ref{Amomdef}), and using the identity
\begin{equation}
\label{BorelFOPT}
-\frac{1}{2\p i}\oint_{|x|=1}\frac{dx}{x}\,x^m\,(-x)^{-\o}=\frac{\sin(\p\o)}{\p(m-\o)}\ ,
\end{equation}
one easily obtains that
\begin{equation}
\label{BorelFOPT2}
A_m(s_0)=-\frac{1}{2\p i}\oint_{|x|=1}\frac{dx}{x}\,x^m\,D(s_0 x)=
\frac{1}{\p}\,\int_0^\infty d\o\,\left(\frac{\sin(\p\o)}{(m-\o)(p-\o)^\g}\right)\,e^{-\frac{\o}{\a(s_0)}}\ .
\end{equation}
The case $\g=1$ is particularly simple as this integral is well-defined for all values of $m$ except $m=p$, for
which it diverges. For $m\neq p$, the integrand in Eq.~(\ref{BorelFOPT2}) is analytic everywhere in the complex $\o$ plane and has an \emph{infinite} radius of convergence in $\o$. This allows the integration to be carried out even after expanding the expression in parenthesis in powers of $\o$, resulting in a convergent power expansion in $\a$ for the integral,  whose radius of convergence we show
to be equal to $1/\p$ in App.~\ref{RoC}. In this case, the Borel transform of FOPT in Eq. (\ref{BorelFOPT2}) defines an analytic function which we will assume  to be the \emph{exact} answer. This will be sufficient for our purposes.\footnote{Even when a convergent power expansion defines an analytic function in $\a$, this function might differ from the exact solution by terms $\mathcal{O}(e^{-1/\a})$. } The situation will be different in the CIPT case, which is key to the
arguments of Ref.~\cite{Hoang:2020mkw} and this paper.

In the general case, the pole at $\o=p$ is not simple but double (\ie, $\g=2$),  or even a branch point if one goes beyond the large-$\b_0$ approximation. In these cases, even FOPT needs to be regulated  and it is common practice to do so according to the Principal Value prescription, defined as the average with respect to an $\pm i\e$ shift in both singularities at $\o=p$ and $\o=m$, taking the limit $\e \to 0$ at the end. The fact that the result has a finite limit when $\e \to 0$ means that we are defining the Borel integral by analytic continuation.

The FOPT series, being an expansion in powers of $\a$,  can therefore be obtained as
\begin{equation}
\label{BorelFOPTSer}
A_m(s_0)=-\frac{1}{2\p i}\oint_{|x|=1}\frac{dx}{x}\,x^m\,D(s_0 x)=
\frac{1}{\p}\,\int_0^\infty d\o\,\left[\frac{\sin(\p\o)}{(m-\o)(p-\o)^\g}\right]_T\,e^{-\frac{\o}{\a(s_0)}}\ ,
\end{equation}
where the symbol $\left[H(\o)\right]_T$ denotes the Taylor expansion of  the function $H(\o)$ around $\o=0$, integrated term by term.

To study the difference between FOPT and CIPT and make contact with the Asymptotic Separation defined in Ref.~\cite{Hoang:2020mkw}, it will prove convenient to split the $\sin(\p\o)=\frac{1}{2i}(e^{ i\p\o}-e^{- i\p\o})$, and
integrate each term in the difference separately. This requires the poles at $\o=p$ and $\o=m$ to be regulated and we will do so using the common Principal Value prescription mentioned above.

\vskip0.8cm
\begin{boldmath}
\section{\label{CIPTAdler} CIPT for the Adler function and arbitrary moment $x^m$ and the Asymptotic Separation of Hoang and Regner}
\end{boldmath}

Unlike in the case of FOPT, the CIPT choice for the scale $\m^2$ in the contour moment (\ref{Amomdef}) corresponds to taking $\mu^2=z$ in Eq.~(\ref{Adler2}). This leads to
\begin{equation}
\label{AdlerCIPT}
D(z)=\int_0^\infty d\o\,\left[\frac{1}{(p-\o)^\g}\right]_T\,e^{-\frac{\o}{\a(-z)}}=\sum_{n=0}^\infty \, n!\, \frac{  B(-\g,n)}{p^{n+\g}}\,\a^{n+1}(-z)\ ,
\end{equation}
where $B(-1,n)=1$, $B(-2,n)=n+1$.

To obtain $A_m(s_0)$ in CIPT, we again use the identity~(\ref{BorelFOPT}) and obtain
\begin{equation}
\label{AmCIPT}
A_m(s_0)=\frac{1}{\p}\int_0^\infty d\o \left[\frac{1}{(p-\o)^\g} \right]_T \frac{\sin(\p \o)}{m-\o} e^{-\o/\a(s_0)}\ ,
\end{equation}
for its corresponding series expansion.    This amounts to carrying out the contour integration of Eq.~(\ref{AdlerCIPT})
term by term, because $\frac{\sin(\p \o)}{m-\o}$ is an analytic function everywhere in the complex plane, and
thus equals its Taylor expansion everywhere.   There is thus no need to deform the contour, as was done
in Ref.~\cite{Hoang:2020mkw}, where the series was Borel resummed first.

We note the difference between the FOPT series in Eq.~(\ref{BorelFOPTSer}) and the CIPT counterpart in Eq.~(\ref{AmCIPT}). Resumming the series in Eq.~(\ref{AmCIPT}) by removing the symbol $\left[\ \right]_T$ immediately reproduces the FOPT result in Eq.\,(\ref{BorelFOPT2}). Why are these two series different then?  In the particular case of a simple pole, $\g=1$,  one can easily see where the difference comes from: The expansion of $\frac{1}{p-\o}$  in powers of $\o$  and consequent integration term by term is a mathematically illegal operation since this expansion converges only
for $|\o|\leq p$, whereas $\o$ is integrated all the way to infinity. This causes the series~(\ref{AmCIPT}) to be asymptotic, unlike the FOPT result in Eq.~(\ref{BorelFOPTSer})  since $\sin(\p\o)/((m-\o)(p-\o))$ has no singularity in $\o$ for $m\ne p$, and  therefore, an infinite radius of convergence in a power expansion in $\o$. This leads to the finite radius of convergence of the
FOPT expansion in powers of $\a$ found in App.~\ref{RoC}.

As an asymptotic expansion, CIPT in Eq.~(\ref{AmCIPT}) will approach a certain value, in general different from Eq.~(\ref{BorelFOPT2}), before it diverges. In Ref.~\cite{Hoang:2020mkw}, a Borel sum for the CIPT expansion
was defined, and an explicit  expression for the difference between the CIPT and FOPT Borel sums, the ``Asymptotic Separation'' (AS) was obtained.

In order to calculate this AS we will follow a slightly different route from Ref.~\cite{Hoang:2020mkw} which, in our opinion, requires less mathematical complexity and may clarify some aspects of the final result, at least in the large-$\b_0$ approximation. To do this, it is useful to split the $\sin(\p \o)$ in Eq.~(\ref{AmCIPT}) into two exponentials and consider the following difference ($\g=1,2$):
\begin{eqnarray}
\label{asympsep}
&&\D(\a,p,m,\g,\e,k)=\\
&&\hspace{-1cm}\frac{1}{2 i\p} \int_0^\infty d\o\,  \frac{1}{(p+ i\e-\o)^\g}\,\ \frac{ e^{-\frac{\o}{\a}\left(1- ik\p\a\right)}}{m+ i \e-\o}- \frac{1}{2 i\p} \int_0^\infty d\o\,  \left[\frac{1}{(p-\o)^\g}\right]_T\,\ \frac{ e^{-\frac{\o}{\a}\left(1- ik\p\a\right)}}{m+ i \e-\o}\ ,\nonumber
\end{eqnarray}
where $\a\equiv\a(s_0)$ and $k=\pm 1$.
Notice how the poles now need to be regulated with the addition of an $i\e$ ($\e>0$), as the $\sin(\p \o)$ function is absent in the numerator. It is intuitively clear that a nonzero difference has to be due to the position of the pole at $\o=p+i\e$ and that, in physics terms, it should be nonperturbative with respect of the coupling $\a$, the expectation being a behavior of the form  $\D\sim \mathcal{O}(e^{-p/\a})$.

To arrive at a result for $\D$ in Eq.~(\ref{asympsep}), we will first study a slightly simpler example. Consider the difference
\begin{equation}
\label{asympseptoy}
\widetilde{\D}=\frac{1}{2 i\p} \int_0^\infty d\o\,  \frac{e^{-\frac{\o}{\a}\left(1- i\p\a\right)}}{p+ i\e-\o}\,- \frac{1}{2 i\p}  \int_0^\infty d\o\,  \left[\frac{1}{p-\o}\right]_T  e^{-\frac{\o}{\a}\left(1- i\p\a\right)}\ .
\end{equation}
The second integral need not be regulated with an $i\e$ because, due to the implicit series expansion in $\o$, the integrand has no pole.

If we could remove the symbol $[\ ]_T$ in the second term of (\ref{asympseptoy}) (which of course we cannot, because of the pole at $\o=p$ which would then appear), the result for the difference would be clearly zero. Instead, before we remove this symbol, we move away from the real axis.

The Taylor expansion of the second term does not contain the pole at $\o=p$ and, consequently, the second integral can actually be taken along any path which does not touch the positive real axis, $\G$, going from $\o=0$ to $\o=\infty$  in the first quadrant of the complex $\o$ plane. For definiteness, we will take $\G$ to be a straight line with a finite angle between $0$ and $\p/2$ with respect to the real axis.  We then have
\begin{equation}
\label{asympseptoy2}
\widetilde{\D}=\frac{1}{2 i\p} \int_0^\infty d\o\,  \frac{e^{-\frac{\o}{\a}\left(1- i\p\a\right)}}{p+ i\e-\o}\,- \frac{1}{2 i\p}\int_\G d\o\,  \left[\frac{1}{p-\o}\right]_T  e^{-\frac{\o}{\a}\left(1- i\p\a\right)}\ .
\end{equation}
If we now define a resummation by removing the symbol $[\ ]_T$, regulating the pole with the same prescription as in the first term, the expression becomes
 \begin{eqnarray}
\label{asympseptoy3}
\widetilde{\D}&=&\frac{1}{2 i\p} \int_0^\infty d\o\,  \frac{e^{-\frac{\o}{\a}\left(1- i\p\a\right)}}{p+ i\e-\o}\,- \frac{1}{2 i\p}  \int_\G d\o\,  \frac{e^{-\frac{\o}{\a}\left(1- i\p\a\right)}}{p+i\e -w}\nonumber\\
&=& \frac{1}{2 i\p} \oint d\o\,  \frac{e^{-\frac{\o}{\a}\left(1- i\p\a\right)}}{p+ i\e-\o}\ ,
\end{eqnarray}
which immediately shows the presence of the pole at $p=\o$ with the help of the the residue theorem (the contribution from the circular arc at infinity connecting the positive real axis and $\G$ vanishes). This procedure defines a Borel sum for the 2nd term and  yields for this difference
\begin{equation}
\label{asympseptoy4}
\widetilde{\D}= (-1)^{p+1}\ e^{-\frac{p}{\a}}\ .
\end{equation}
As we will see below, this result for $ \widetilde{\D}$ previews the AS defined in Ref.~\cite{Hoang:2020mkw} for this case of Eq.~(\ref{asympseptoy}).
%%%%%%%%%%%%%%%%%%%
\begin{figure}
\vspace*{4ex}
\begin{center}
\includegraphics*[width=7cm]{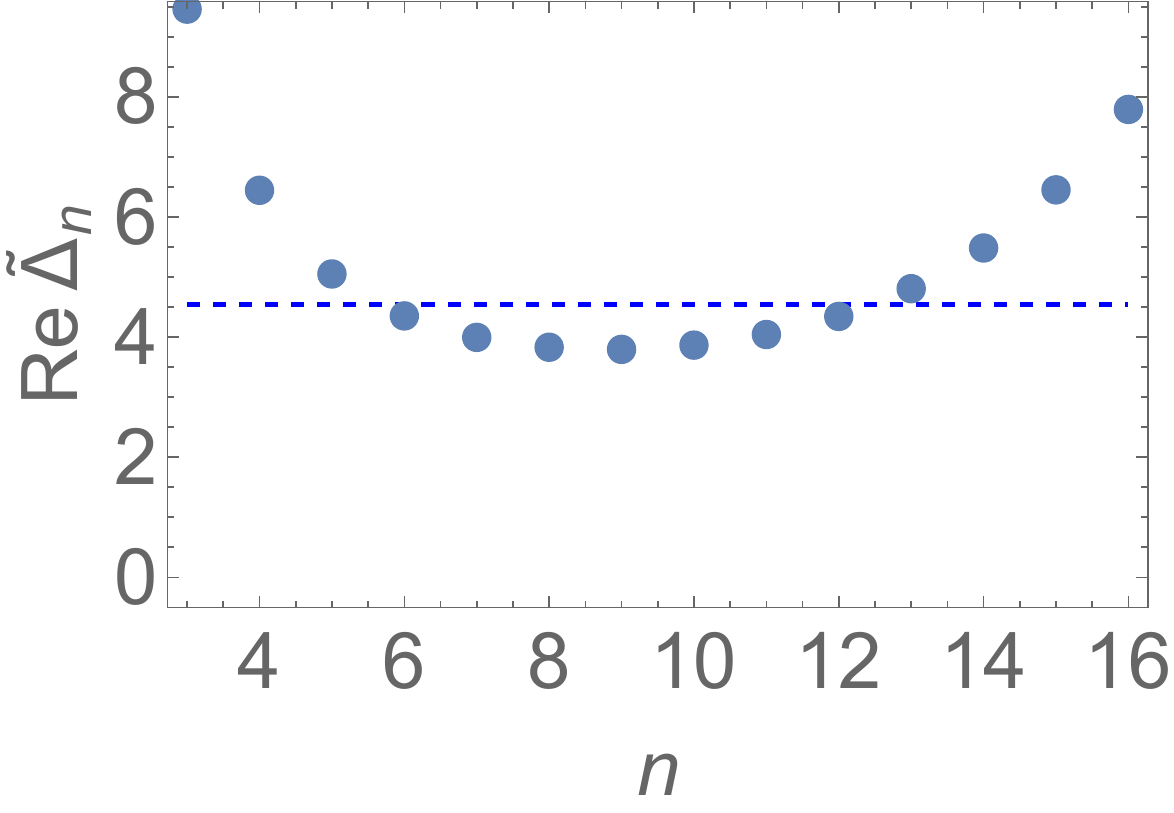}
\hspace{.5cm}
\includegraphics*[width=7cm]{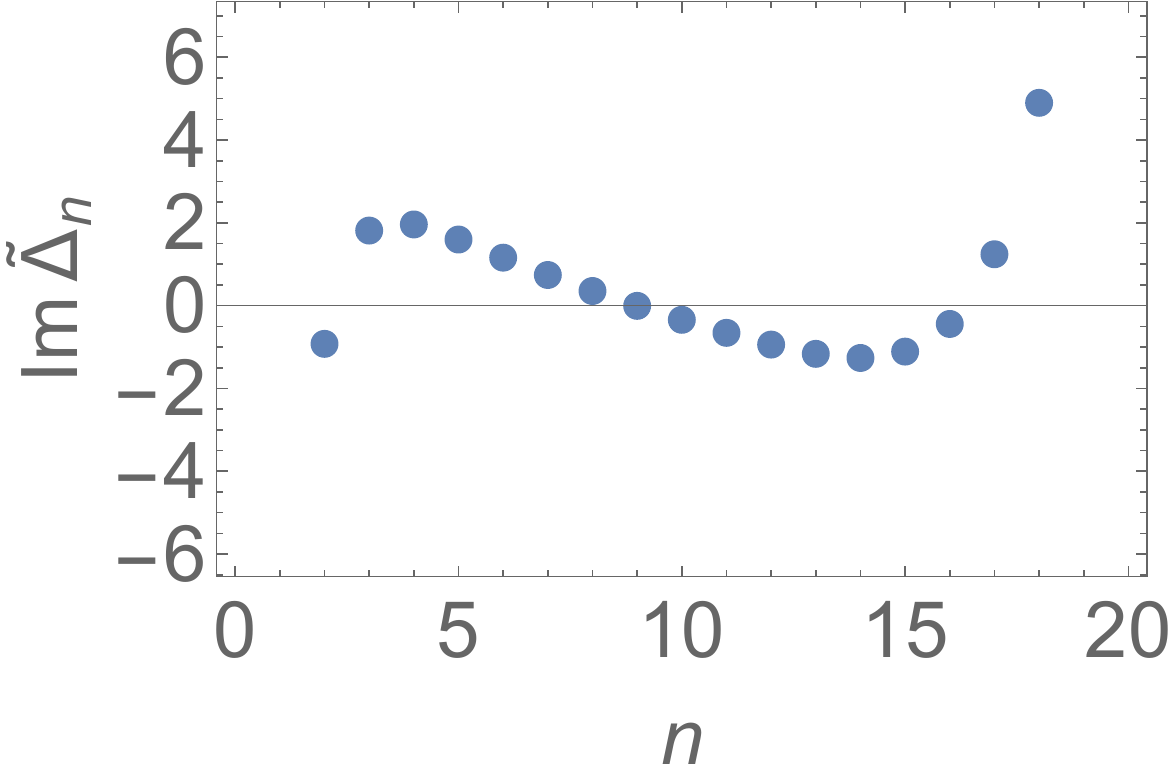}
\end{center}
\begin{quotation}
\floatcaption{toyfigs}%
{{\it Left panel: Real part of Eq.~(\ref{Deltatilden}) for $\a=1/10$ and $p=1$, as a function of the order, $n$, multiplied by $10^5$. The dashed line is the result of  $\widetilde{\D}$, Eq.~(\ref{asympseptoy4}). Right panel: same for the imaginary part, with} $\mathrm{Im}\ \widetilde{\D}=0$.}
\end{quotation}
\vspace*{-4ex}
\end{figure}
%%%%%%%%%%%%%%%%%%%
Figure~\ref{toyfigs} shows
\begin{equation}
\label{Deltatilden}
\widetilde{\D}_n=\frac{1}{2 i\p} \int_0^\infty d\o\,  \frac{e^{-\frac{\o}{\a}\left(1- i\p\a\right)}}{p+ i\e-\o}\,- \frac{1}{2 i\p p}  \int_0^\infty d\o\,  \sum_{k=0}^n\left(\frac{\o}{p}\right)^k  e^{-\frac{\o}{\a}\left(1- i\p\a\right)}
\end{equation}
as a function of $n$.  The resummed $\widetilde{\D}$ of Eq.~(\ref{asympseptoy3}) is purely real even though both integrals in Eq.~(\ref{asympseptoy3}) are complex. This means that the real and imaginary parts of  Eq.~(\ref{Deltatilden}) behave very differently. Indeed one can see on the left panel that the real part approaches a nonzero value, given approximately by
 the expression in Eq.~(\ref{asympseptoy4}) before it diverges,
 whereas the imaginary part in the right panel oscillates around zero before diverging. If we had chosen to regulate the pole with $-i\e$ instead, the pole at $\o=p-i\e$ would not be encircled by the contour in Eq.~(\ref{asympseptoy3}) and the result would be zero. This choice of $i\e$ prescription would lead to a plot like that in the right panel of Fig.~\ref{toyfigs}, but now for both the real and the imaginary part: The series oscillates around the result and there is no asymptotic separation.\footnote{The behavior shown in the right panel of Fig.~\ref{toyfigs} is reminiscent of the case where $p$ (the position of the pole in the Adler function) is the same as $m$ (the position of the pole generated upon contour integration of the moment). See section \ref{pNOTm} below.}

We are now ready to calculate the AS for the difference of FOPT in Eq.~(\ref{BorelFOPT2}) and CIPT in Eq.~(\ref{AmCIPT}). To this end, it is convenient to define the integrals\footnote{As before, we choose $\e>0$ in what follows but choosing $\e<0$ leads to the same result.}
\begin{equation}
\label{master}
I(\a,p,m,\g,\e,k)=\frac{1}{2 i \p}\int_0^\infty d\o\, \frac{1}{(p+i\e-\o)^\g}\, \frac{e^{-\frac{\o}{\a}(1- \, i\p\a\, k)}}{m+i\e-\o}\ ,
\end{equation}
and
\begin{equation}
\label{masterT}
I_T(\a,p,m,\g,\e,k)=\frac{1}{2 i \p}\int_0^\infty d\o \left[\frac{1}{(p+i\e-\o)^\g}\right]_T\, \frac{e^{-\frac{\o}{\a}(1- \, i\p\a\, k)}}{m+i\e-\o} \ .
\end{equation}
The result for the FOPT moment $A_m(s_0)$ in Eq.~(\ref{BorelFOPT2}) may be expressed as
\begin{eqnarray}
\label{FOPTmaster}
 A^{\rm FOPT}_m(s_0)&=&\frac{1}{2}\Big[I(\a,p,m,\g,\e,k=1)+ I(\a,p,m,\g,-\e,k=1)  \\
 &&-I(\a,p,m,\g,\e,k=-1)-I(\a,p,m,\g,-\e,k=-1)\Big]\ ,\nonumber
 \end{eqnarray}
 where the $\e\rightarrow 0$ limit at the end is understood.\footnote{When $\g=1$ this limit is trivial due to the $\sin(\p\o)$ function in Eq.~(\ref{BorelFOPT2}).} The equivalent CIPT version, $A^{\rm CIPT}_m(s_0)$ is obtained from the above expression using the integral $I_T$ instead of $I$.
Repeating the same steps which led to the result of $\widetilde{\D}$ in Eq.~(\ref{asympseptoy3}), we get ($\e>0$):
\begin{equation}
\label{ASlargebeta}
\D_{\rm AS}(\a,p,m,\g)=\frac{1}{2}\Big[\D(\e,1)+\D(-\e,1)-\D(\e,-1)-\D(-\e,-1)\Big]\ ,
\end{equation}
where for simplicity we omit the arguments $\a$, $p$, $m$ and $\g$ and
\begin{eqnarray}
\label{delta}
\D(\e,k)&=&\frac{1}{2i\p}\int_{0}^{\infty}d\o \frac{1}{(p+i\e-\o)^\g}\, \frac{e^{-\frac{\o}{\a}(1-\, i\p\a\, k)}}{m+i\e-\o}\\
&&-\frac{1}{2i\p}\int_{\G_k}d\o \frac{1}{(p+i\e-\o)^\g}\, \frac{e^{-\frac{\o}{\a}(1- \, i\p\a\, k)}}{m+i\e-\o} \nonumber \\
&&-k\ \Theta(k\e)\, \mathrm{Res}\left\{ \frac{1}{(p+i\e-\o)^\g}\, \frac{e^{-\frac{\o}{\a}(1-\, i\p\a\, k)}}{m+i\e-\o}; \o=m+i\e \right\} \nonumber\\
&=& k\ \Theta(k\e)\, \mathrm{Res}\left\{ \frac{1}{(p+i\e-\o)^\g}\, \frac{e^{-\frac{\o}{\a}(1-\, i\p\a\, k)}}{m+i\e-\o}; \o=p+i\e \right\}\ .
\end{eqnarray}
Here $\G_k$ is a straight line from $\o=0$ to $\o=\infty$ in the first quadrant if $k=1$ and in the fourth quadrant if $k=-1$. The function $\Theta(x)=1$ for $x>0$ and $\Theta(x)=0$ for $x<0$, which implies that $\D(-\e,1)=\D(\e,-1)=0$ because no pole is encircled in these cases. Therefore, Eq.~(\ref{ASlargebeta})
simplifies to the sum of two  separate contour  integrals along two closed contours (one counter-clockwise in the first quadrant when $k=1$ and one clockwise in the fourth quadrant when $k=-1$) and we find as the final result
\begin{equation}
\label{ASlargebetaResult}
\D_{\rm AS}(\a,p,m,\g)=\, \mathrm{Res}\left\{ \frac{1}{(p-\o)^\g}\, \frac{e^{-\frac{\o}{\a}} \cos(\p \o)}{m-\o}; \o=p\right\}\ ,
\end{equation}
where we have already taken the limit $\e\to 0$. This result immediately shows why the IR renormalons are the ones responsible for the AS. UV renormalons, on the other hand,  do not contribute as their poles are located outside of  both integration contours and, correspondingly, their residues vanish.

We may now list a couple of explicit results:
\begin{eqnarray}
\label{resultAS}
&&\D_{\rm AS}(\a,p,m,1)=(-1)^p\, \frac{e^{-p/\a}}{p-m} \ ,\\
&&\D_{\rm AS}(\a,p,m,2)=(-1)^p\, \ e^{-p/\a}\left(\frac{1}{(p-m)^2}+ \frac{1}{\a}\,\frac{1}{p-m} \right)\ ,\nonumber
\end{eqnarray}
in agreement with the results found in Ref.~\cite{Hoang:2020mkw}.\footnote{Note that Ref.~\cite{Hoang:2020mkw} defines the moments with $(-x)^m$, so that their expressions differ
by $(-1)^m$ from ours.} This shows the equivalence of both methods, at least for the large-$\b_0$ approximation considered here. Clearly, the case $m=p$ is special as the above expressions for $\D_{\rm AS}(\a,p,p,\g)$ are ill-defined.  In this case, Hoang and Regner in Ref.~\cite{Hoang:2020mkw} choose to define $\D_{\rm AS}$ to be zero because the CIPT series shows in this case a behavior similar to that in the right panel of Fig.~\ref{toyfigs}. This is a situation  where CIPT crosses the FOPT result between two consecutive orders in the expansion, even though in general there is no single order for which the CIPT series agrees with the FOPT result.

\vskip0.8cm
\begin{boldmath}
\section{\label{Minimum Distance} The Minimum Distance (MD) between the FOPT and the CIPT Series}
\end{boldmath}

In this section we would like to discuss another possible approach to the FOPT-CIPT difference which, unlike that of Ref.~\cite{Hoang:2020mkw}, is based on the CIPT series for the contour moments $A_m(s_0)$ themselves rather than on its Borel resummation. After all, it is the series one considers in the actual analyses of
hadronic $\tau$ decays in practice.

The starting point of this discussion is the fact that the difference between the FOPT Borel representation in Eq.~(\ref{BorelFOPT2}) and the sum of the first $n-1$ terms of the CIPT \emph{series expansion} in Eq.~(\ref{AmCIPT}) can be written as ($\g=1,2$)
\begin{equation}
\label{diffser}
\D_{\rm SER}(\a,p,m,\g,n)=\frac{1}{\p}\int_0^\infty \!\!\! d\o\, \frac{e^{-\o/\a}\ \sin(\p \o)}{(m-\o+i\e)(p-\o+i\e)^\g}\left(\frac{\o}{p}\right)^n \left(\frac{p+n(p-\o)}{p}\right)^{\g-1}\ ,
\end{equation}
where the limit $\e\to 0$ will always be taken in the end. The series $\D_{\rm SER}(\a,p,m,\g,n)$ constitutes an asymptotic series as a function of the order $n$ and, as such, it comes within a  minimum distance of zero for some value of $n=n^*$, where it stabilizes for a few orders before it diverges. The value of this minimum distance at the order $n=n^*$ is the difference  between the FOPT Borel result and the CIPT series, $\D_{\rm SER}(\a,p,m,\g,n^*)$, we are interested in. General properties of asymptotic expansions suggest that we should expect this value of $n^*$ to be of $\mathcal{O}(1/\a)$, and this minimum distance of $\mathcal{O}(e^{-p/\a})$.

In order to verify that this is indeed the case, we carry out a saddle point approximation and express $\D_{\rm SER}$ as
\begin{equation}
\label{diffserSaddle}
\D_{\rm SER}(\a,p,m,\g,n)=\frac{1}{\p}\int_0^\infty  d\o\, \frac{e^{\Phi(\o,n)}\ \sin(\p \o)}{(m-\o+i\e)(p-\o+i\e)^\g} \left(\frac{p+n(p-\o)}{p}\right)^{\g-1}\ ,
\end{equation}
where
\begin{eqnarray}
\label{phi}
\Phi(\o,n)&=&-\frac{\o}{\a}+ n \log\frac{\o}{p}\nonumber\\
&\approx& -\frac{\o_0}{\a}+ n \log\frac{\o_0}{p}- \frac{n}{2}\frac{(\o-\o_0)^2}{{\o_0}^2}\ ,
\end{eqnarray}
and $\o_0=n \a$ is the value at which the derivative vanishes, $\Phi'(\o_0,n)=0$.
Setting $\o_0=n\a$, and
imposing the asymptotic stability condition $\Phi(\o,n^*)\sim \Phi(\o,n^*+1)$ leads to $n^*=p/\a$ and, consequently, $\o_0=p$.\footnote{The factor $\frac{p+n(p-\o)}{p}$ is equal to one for $\o=\o_0=p$.} The exponential factor in (\ref{diffserSaddle}), therefore, becomes
\begin{equation}
\label{expsaddle}
e^{\Phi(\o,n^*)}\approx e^{-\frac{p}{\a}}\  e^{-\frac{1}{2}\, \frac{(\o-p)^2}{p\a}}\quad ,\quad n^*= \frac{p}{\a}\ ,
\end{equation}
which yields a nonperturbative correction $\mathcal{O}( e^{-\frac{p}{\a}})$, as expected, times a gaussian centered at $\o=p$ with a narrow width of $ \mathcal{O}( \sqrt{p\, \a})$. In fact, for $\a$ small enough, the gaussian is so narrow that, when the integrand in Eq.~(\ref{diffserSaddle}) is well defined at $\o=p$ (as it is the case for $\g=1$ and $m\neq p$), this gaussian may be replaced by a Dirac delta, \ie,  $\sim \d(\o-p)$. In particular when $p=-|p|$ is a negative integer, as it is the case of UV renormalons, the $i\e$ prescription is not necessary since the integrand is regulated by the $\sin(\p \o)$ and the integral~(\ref{diffser}) vanishes like $\sin(\p \o)\, \d(\o-|p|)/(p-\o)$.
 If the gaussian is not approximated by the Dirac delta,  it will still be very much suppressed relative to the natural scale $\mathcal{O}(e^{-|p|/\a})$. The net result is that UV renormalons are not the main cause for the FOPT-CIPT difference, in agreement with the conclusion based on the AS criterion of Ref.~\cite{Hoang:2020mkw} and our discussion in the previous section.

We have checked the results~(\ref{expsaddle}) in a number of cases and the agreement for small $\a$ is very good, being optimal  when the value of $n^*=p/\a$ is employed in the initial integral $\D_{\rm SER}$ in Eq.~(\ref{diffser}) rather than using the alternate expression
obtained using the approximation (\ref{phi}) in Eq.~(\ref{diffserSaddle}). A possible exception appears to be the double pole in the Adler function for the case with  $m=p$, for which, as we will see below,  the subleading terms in $n^*=p/\a+ ...$ are not negligible.

In what follows, we will compare in a few cases the ``Minimum Distance" (MD), defined as
$\D_{\rm SER}(\a,p,m,\g,n^*)$ with $n^*=p/\a$, with
$\D_{\rm SER}(\a,p,m,\g,n)$
as a function of $n$, and also with the corresponding result for the AS of Ref.~\cite{Hoang:2020mkw}.

\vskip0.8cm
%\newpage
\begin{boldmath}
\subsection{\label{pNOTm} The case $p\neq m$}
\end{boldmath}
An interesting case in this category is the case $p=2$ and $\g=1$, since that respresents the simple pole corresponding to the first IR renormalon \cite{Broadhurst:1992si} (see also Ref.~\cite{Beneke:1998ui}). In this case we can easily obtain an analytic expression for the MD with the help of the Dirac delta approximation for the exponential in Eq.~(\ref{expsaddle}),
\begin{equation}
\label{diracdelta}
e^{-\frac{1}{2}\, \frac{(\o-p)^2}{p\a}} \approx \sqrt{2 p\p \a}\, \d(\o-p) \quad, \quad (\a \to 0)\ ,
\end{equation}
which, when inserted into Eq.~(\ref{diffserSaddle}), yields for the MD
\begin{equation}
\label{MDanalytic}
 \D_{\rm SER}(\a,p,m,1,n^*)\approx (-1)^p \sqrt{2 p\p \a}\, \frac{e^{-p/\a}}{p-m}\quad,\quad (\a \to 0)\ .
\end{equation}
This result should be compared to the expression for the AS in the first of Eq.~(\ref{resultAS}).

In general, however, instead of the Dirac delta approximation, it is better to use the value of $n^*=p/\a$ in the expression~(\ref{diffserSaddle}).   Choosing $m=1$ for example, Fig.~\ref{g1p2m1} shows a comparison of the MD with the result for $\D_{\rm SER}(\a,p,m,\g,n)$ as a function of $n$ for $\a=1/5$ and for $\a=1/40$. From the figure one sees that the AS of Ref.~\cite{Hoang:2020mkw} does well for a value of $\a=1/5$, as the MD does too, but for a value of $\a$ much smaller, such as \eg\, $\a=1/40$ (deeper in the asymptotic regime), the  MD gives a better description of the FOPT-CIPT difference.

%%%%%%%%%%%%%%%%%%%
\begin{figure}
\vspace*{4ex}
\begin{center}
\includegraphics*[width=7cm]{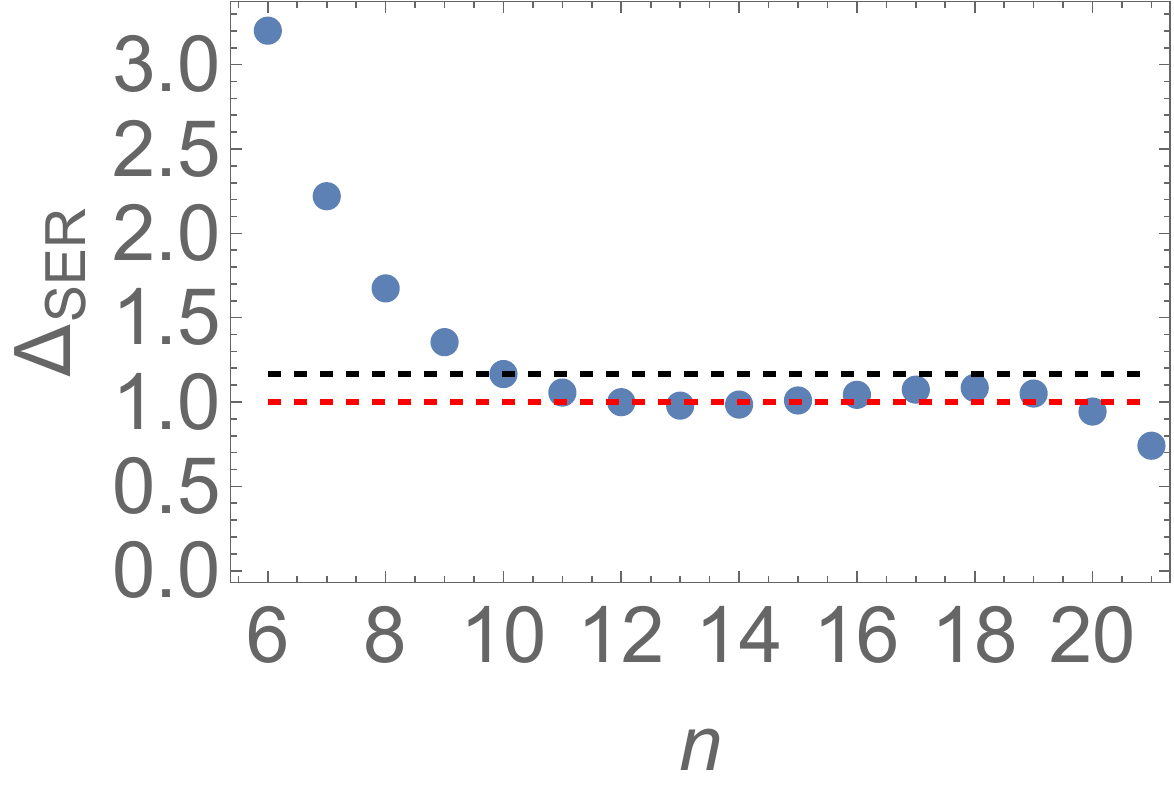}
\hspace{.5cm}
\includegraphics*[width=7cm]{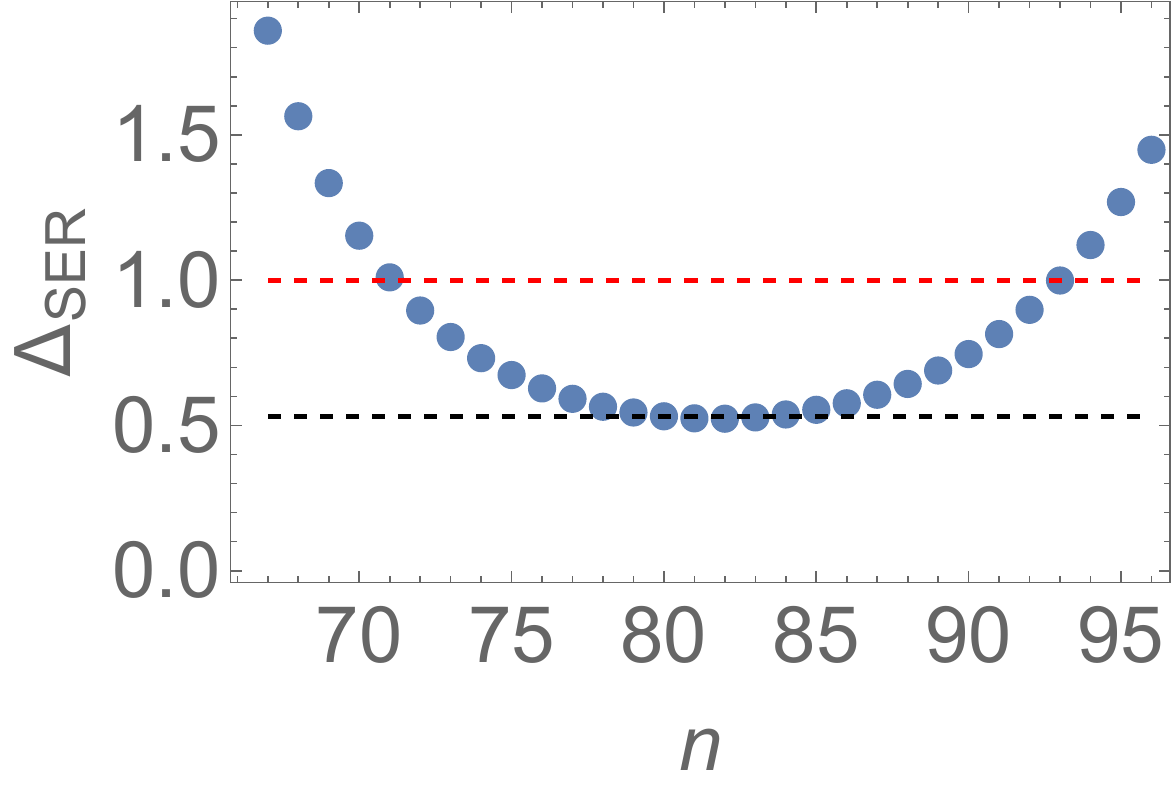}
\end{center}
\begin{quotation}
\floatcaption{g1p2m1}%
{{\it Left panel: Comparison between $\D_{\rm SER}(1/5,2,1,1,n)$ in units of $e^{-\frac{p}{\a}}(=e^{-10}$) as a function of $n$ (blue dots), the MD, \ie, the same function evaluated at $n^*=p/\a$ (black dashed), and the AS result of Ref.~\cite{Hoang:2020mkw} (red dashed) . Right panel: same for $\a=1/40$ in units of $e^{-\frac{p}{\a}}=e^{-80}$.}}
\end{quotation}
\vspace*{-4ex}
\end{figure}
%%%%%%%%%%%%%%%%%%%
%%%%%%%%%%%%%%%%%%%
\begin{figure}
\vspace*{4ex}
\begin{center}
\includegraphics*[width=7cm]{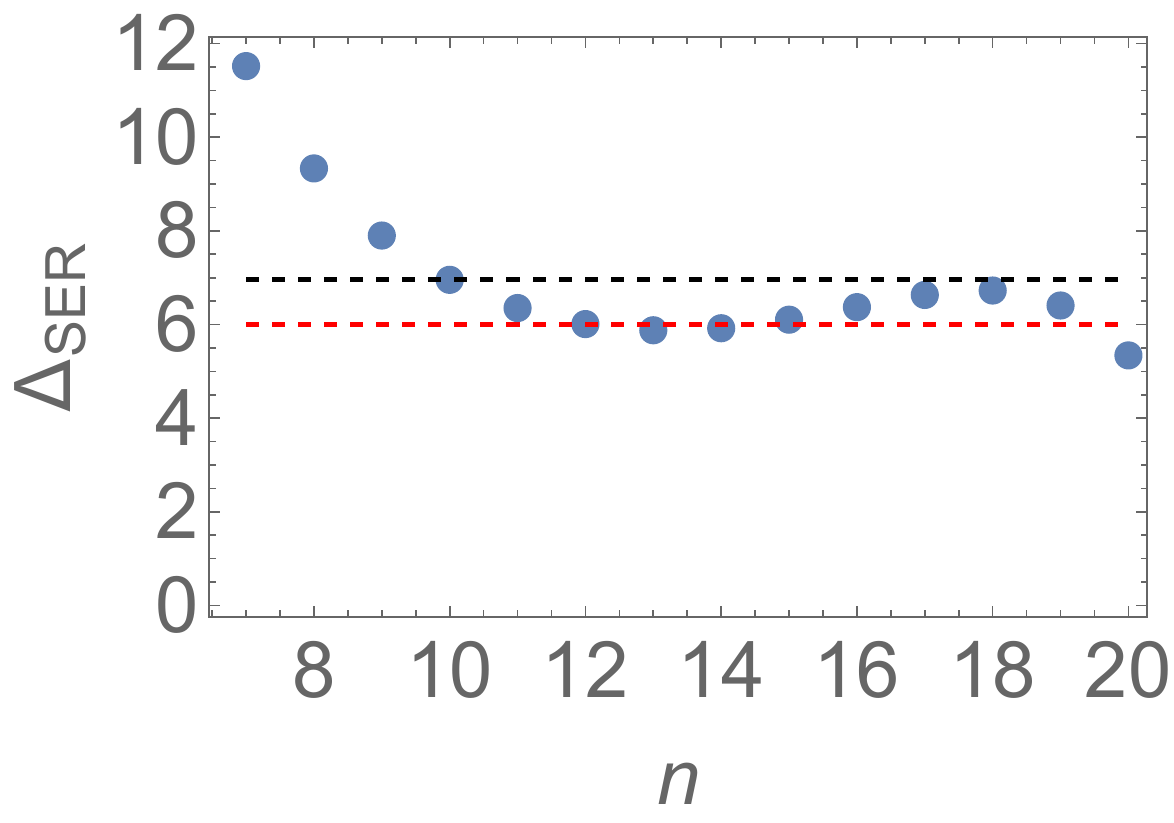}
\hspace{.5cm}
\includegraphics*[width=7cm]{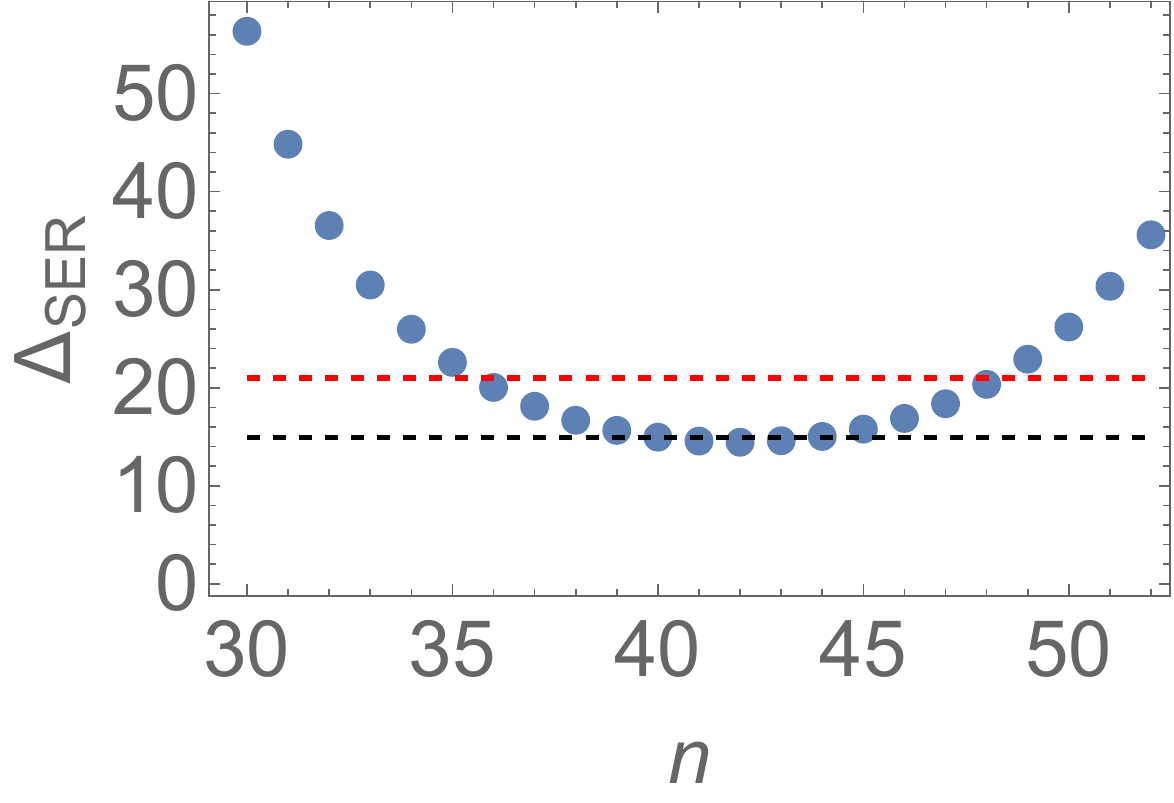}
\end{center}
\begin{quotation}
\floatcaption{g2p2m1}%
{{\it Left panel: Comparison between $\D_{\rm SER}(1/5,2,1,2,n)$ in units of $e^{-\frac{p}{\a}}(=e^{-10})$ as a function of $n$ (blue dots), the MD, \ie, the same function evaluated at $n^*=p/\a$ (black dashed), and the AS result of Ref.~\cite{Hoang:2020mkw} (red dashed). Right panel: same for $\a=1/20$ in units of $e^{-\frac{p}{\a}}(=e^{-40})$.}}
\end{quotation}
\vspace*{-4ex}
\end{figure}
%%%%%%%%%%%%%%%%%%%

Double poles in the Adler function at large-$\b_0$ start at $p=3$ for which the exponential suppression in Eq.~(\ref{expsaddle}) makes the contribution very small. For illustration, we may instead look at a double pole at $p=2$. Again, choosing $m=1$, Fig.~\ref{g2p2m1} shows the result of this comparison. For $\a=1/5$ the AS and the MD fare well but for $\a=1/20$ the MD again describes the FOPT-CIPT difference better.

\vskip0.8cm
\begin{boldmath}
\subsection{\label{pEQUALSm} The special case $p= m$}
\end{boldmath}
This case requires special attention as the result for the AS~(\ref{resultAS})  is ill-defined for $p=m$. In Ref.~\cite{Hoang:2020mkw} the authors define the FOPT-CIPT asymptotic separation in this case to be zero. Although it is true that the difference $\D_{\rm SER}(\a,p,p,\g,n)$ crosses zero at some  intermediate real number between two consecutive integers, $n$ and $n+1$, in general this intermediate value does not coincide with any integer $n$ and, therefore, there is no order in the expansion where one would really expect to find a vanishing result. A similar thing happens with the well-known Stieltjes asymptotic expansion
\begin{equation}
\label{Euler}
I(\a)=\int_0^\infty dt\,  \frac{e^{-t/\a}}{1+t}\approx \sum_{n=0}^N (-1)^n n!\ \a^{n+1}
\end{equation}
which crisscrosses the exact result for this integral at multiple values of $N$ around the asymptotic value $N^*=1/\a$. However, the minimum distance between the series and the exact integral is nonzero but
\begin{equation}
\sqrt{\frac{\p \a}{2}}\, e^{-1/\a}\ ,
\end{equation}
as is well known \cite{Bender}.

In Fig.~\ref{g1p2m2} we show the comparison for the simple pole at $p=m=2$ between the absolute value  $|\D_{\rm SER}(\a,2,2,1,n)|$ and the corresponding MD at $n^*=2/\a$ for two values of $\a$, $\a=1/5$ and $\a=1/30$. Showing the absolute value of $\D_{\rm SER}$ emphasizes the point at which the distance between FOPT and CIPT is a minimum. Although the MD at $n^*$ is small on the scale of $e^{-p/\a}$, it  is not zero.

Finally, in Fig.~\ref{g2p2m2} we show similar results for a double pole at $p=m=2$, $\D_{\rm SER}(\a,2,2,2,n)$, again in absolute value, and for $\a=1/5$ and $\a=1/30$. Again, the minimum value of $|\D_{\rm SER}|$ is nonzero. However, as the figure shows, in this case the asymptotic estimate $n^*=p/\a$ is not sufficiently accurate to exactly reproduce the minimum value of $\D_{\rm SER}$ represented by the lower points. We have checked that using the value $n^*=p/\a-1$, instead, does reproduce the minimum shown in the right panel of Fig.~\ref{g2p2m2} accurately.\footnote{The left panel is for a too small value of $\a$ for the asymptotic estimate to work.} So this points to a situation where the subleading terms in $n^*=p/\a+...$ are not negligible.

%%%%%%%%%%%%%%%%%%%
\begin{figure}
\vspace*{4ex}
\begin{center}
\includegraphics*[width=7cm]{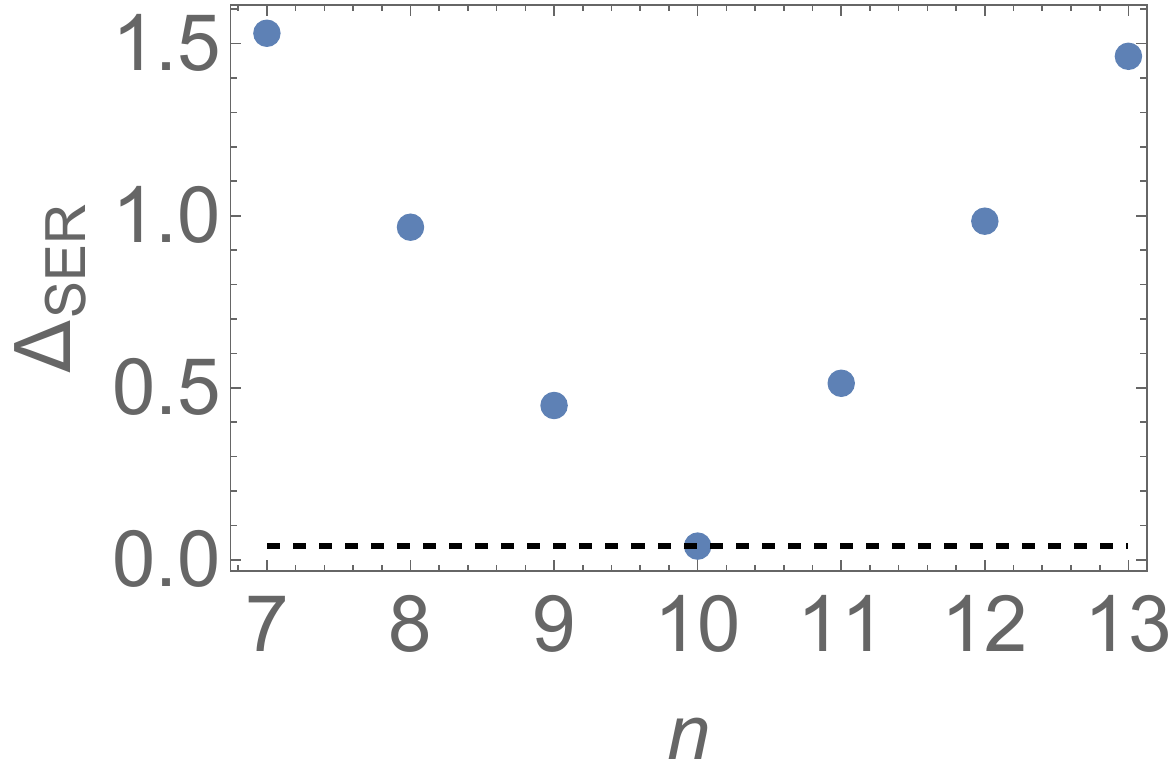}
\hspace{.5cm}
\includegraphics*[width=7cm]{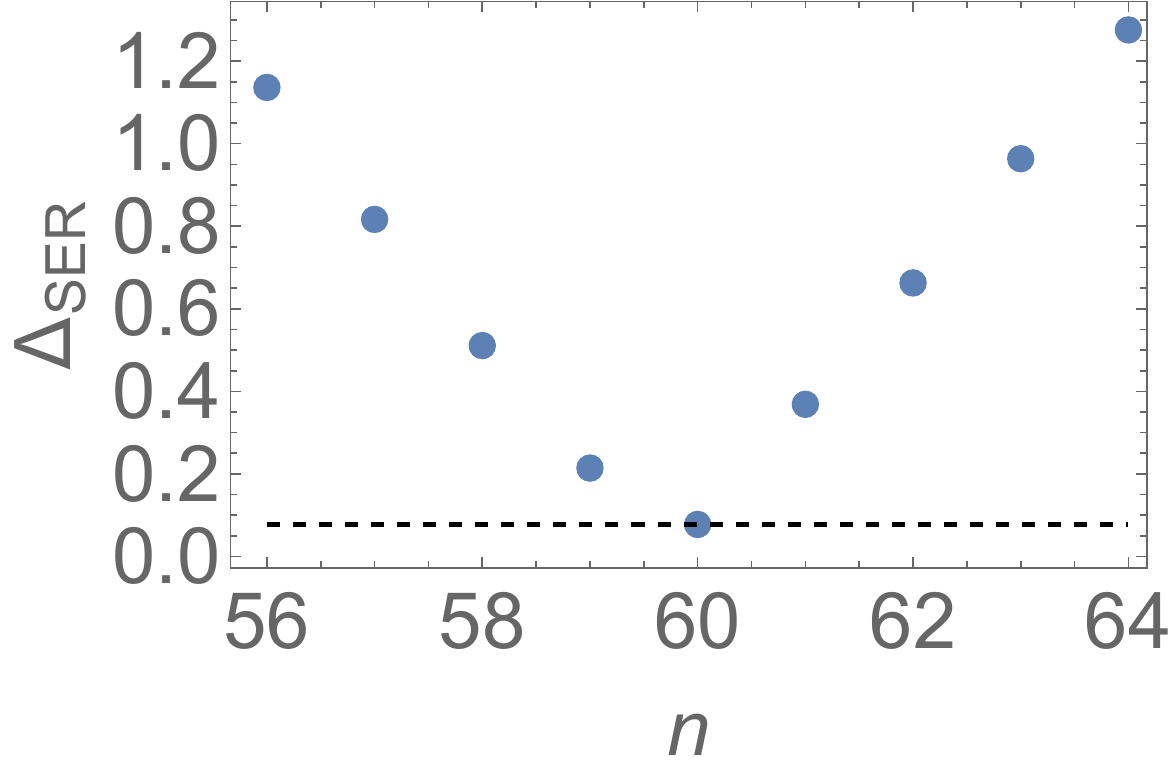}
\end{center}
\begin{quotation}
\floatcaption{g1p2m2}%
{{\it Left panel: Comparison between $|\D_{\rm SER}(1/5,2,2,1,n)|$ in units of $e^{-\frac{p}{\a}}(=e^{-10})$ as a function of $n$ (blue dots) and the MD, \ie, the same function evaluated at $n^*=p/\a$ (black dashed). Right panel: same for $\a=1/30$ in units of $e^{-\frac{p}{\a}}(=e^{-60})$.}}
\end{quotation}
\vspace*{-4ex}
\end{figure}
%%%%%%%%%%%%%%%%%%%

%%%%%%%%%%%%%%%%%%%
\begin{figure}
\vspace*{4ex}
\begin{center}
\includegraphics*[width=7cm]{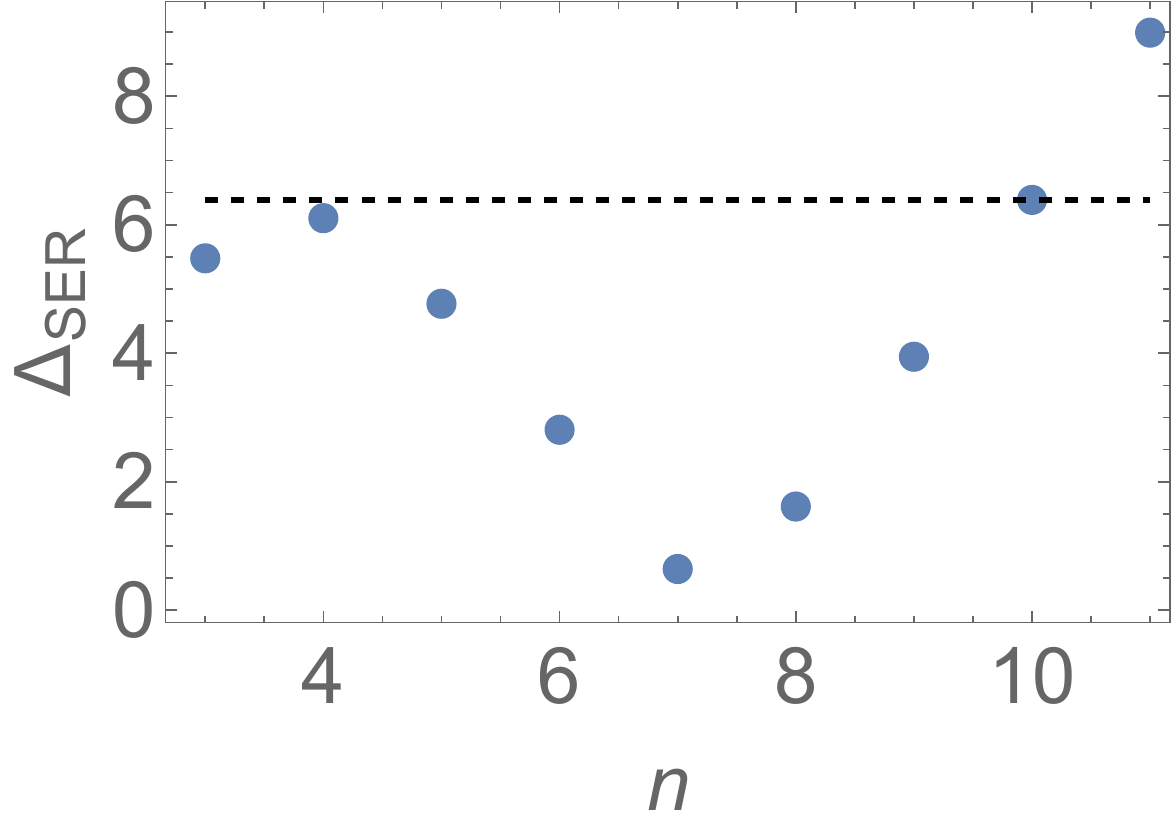}
\hspace{.5cm}
\includegraphics*[width=7cm]{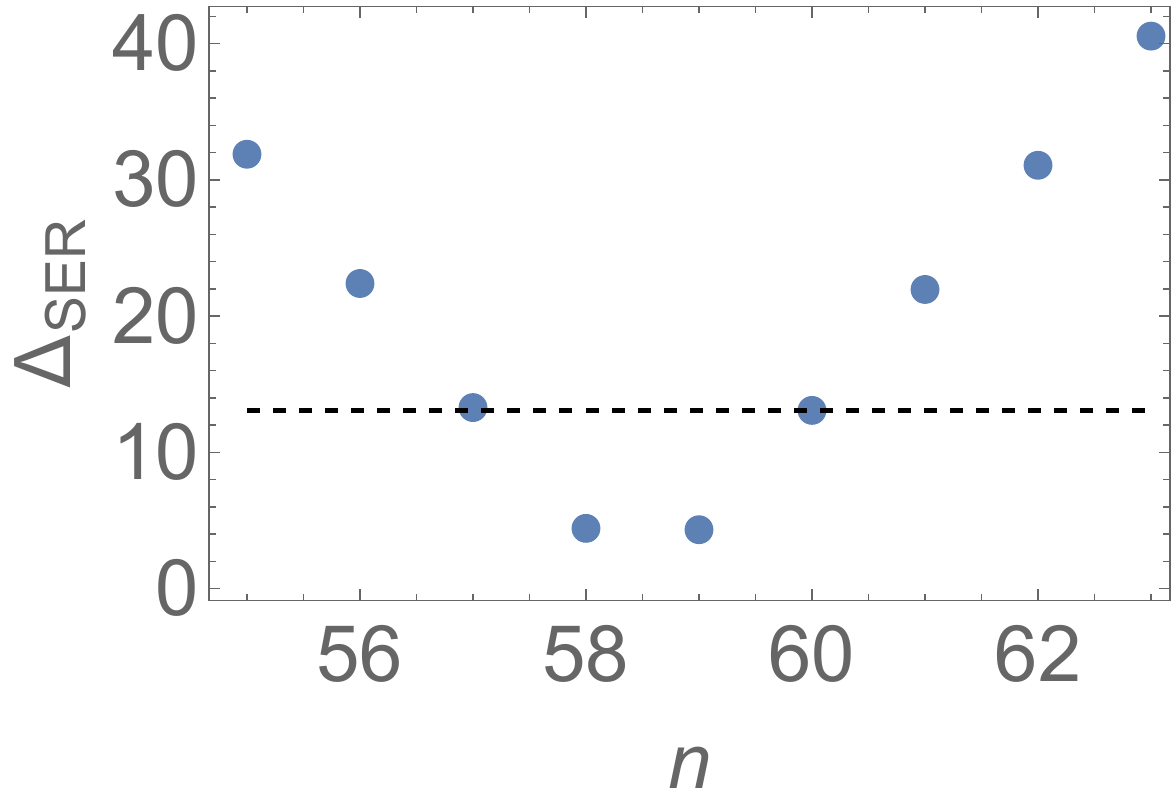}
\end{center}
\begin{quotation}
\floatcaption{g2p2m2}%
{{\it Left panel: Comparison between $|\D_{\rm SER}(1/5,2,2,2,n)|$ in units of $e^{-\frac{p}{\a}}(=e^{-10})$ as a function of $n$ (blue dots), and the MD, \ie, the same function evaluated at $n^*=p/\a$ (black dashed). Right panel: same for $\a=1/30$ in units of $e^{-\frac{p}{\a}}(=e^{-60})$.}}
\end{quotation}
\vspace*{-4ex}
\end{figure}
%%%%%%%%%%%%%%%%%%%

\vskip0.8cm
\begin{boldmath}
\section{\label{conclusions} Conclusions}
\end{boldmath}

For many years the difference between FOPT and CIPT has been a source of systematic uncertainty that has plagued the determination of $\a_s$ from hadronic $\t$ decay. Recently, the authors of Ref.~\cite{Hoang:2020mkw} have pointed out that there is a measure for this difference in terms of the corresponding Borel transforms, which they termed the Asymptotic Separation. This separation is nonperturbative, of $\mathcal{O}(e^{-1/\a_s})$, and therefore in conflict with the OPE, which  makes  CIPT-based results for $\a_s$ inconsistent with the OPE. For example, for $m=1$ the results in Eqs.~(\ref{resultAS}) and~(\ref{MDanalytic}) show a nonperturbative result which has no OPE counterpart as the residue theorem  removes any power corrections to the corresponding moment.\footnote{This is true in the large-$\b_0$ approximation but there is no reason to believe this does not carry over to the case of full QCD.} Consequently, CIPT results should not be included in any average for $\a_s$, at least until schemes such as that proposed in Ref.~\cite{Benitez-Rathgeb:2022yqb}, designed to fix the problems of CIPT, lead to new determinations of $\a_s$ consistent with the OPE.

In the present work, using the large-$\b_0$ approximation, we have confirmed the findings in Ref.~\cite{Hoang:2020mkw}. Using standard analysis tools for asymptotic series, we have introduced a measure of the distance between the FOPT Borel sum and the CIPT perturbative series directly, and not in terms of a  Borel sum for the CIPT series, which we called the Minimum Distance. As in the case of the AS, we have found several examples where this Minimum Distance is not zero, and we believe this result to be generic. In the large-$\b_0$ approximation employed here, there are cases (\eg\  the simple pole, $\g=1$ and $m\ne p$) where the FOPT series is convergent to its Borel sum and, consequently, this sum can be taken to be the exact answer for this series.

The Minimum Distance characterizes the nonperturbative difference between the FOPT and CIPT series without the
need to define a Borel sum for the CIPT series.   As such, unlike the AS, it relies on defining a Borel sum only
for the FOPT series, which is consistent with the OPE.   It thus reinforces the conclusion that the CIPT
series is inconsistent with the OPE.

\vspace{1cm}
\noindent
{\bf Acknowledgements}\\

We thank Diogo Boito for discussions and comments on the manuscript.
The work of MG is  supported
by the U.S. Department of Energy, Office of Science, Office of High Energy
Physics, under Award No. DE-SC0013682.   KM is supported by a grant from the Natural Sciences
and Engineering Research  Council of Canada. SP is supported by the
Spanish Ministry of Science, Innovation and Universities (project
PID2020-112965GB-I00/AEI/10.13039/501100011033)  and by Grant 2017 SGR 1069.
IFAE is partially funded by the CERCA program of the Generalitat de Catalunya.

%\newpage

%%%%%%%%%%%%%%%%%%%%%%%%%%%%%%%%%%%%%%%%
\vskip0.8cm
\appendix
\begin{boldmath}
\section{\label{RoC} Convergence of the FOPT series for $A_m(s_0)$ in $\a$ for $\g=1$ and $p \ne m$}
\end{boldmath}

Here we consider the FOPT series (\ref{BorelFOPTSer}) for $\g=1$ and $p\ne m$:
\begin{equation}
\label{FOPTseriesagain}
A_m(s_0)=\frac{1}{\p}\,\int_0^\infty d\o\,\left[\frac{\sin(\p\o)}{(m-\o)(p-\o)^\g}\right]_T\,e^{-\frac{\o}{\a(s_0)}}\ , \nonumber
\end{equation}
which can be rewritten as
\begin{eqnarray}
\label{rewrite}
A_m(s_0)&=&\frac{1}{\p(m-p)}\left(I(s_0,p)-I(s_0,m)\right)\ ,\\
I(s_0,p)&=&
\int_0^\infty d\o\,\left[\frac{ \sin(\p\o)}{p-\o}\right]_T\,e^{-\frac{\o}{\a(s_0)}}\ .\nonumber
\end{eqnarray}
For $m\ne p$ there can be no cancellation between the two terms. So, to obtain the radius of convergence, we can study just one of them. Let us take the first integral and expand the $\sin(\p \o)$ around $\o=p$:
\begin{eqnarray}
\label{sineexpansion}
\sin(\p\o)&=&\sum_{n\geq1}\frac{(-1)^p \p^n \sin(n \p/2)}{n!}\, (\o-p)^n \\
&=&(-1)^p \p (\o-p)+ \sum_{n\geq 3}\frac{(-1)^p \p^n \sin(n \p/2)}{n!}\, (\o-p)^n \nonumber\ ,
\end{eqnarray}
and insert this identity into $I(s_0,p)$ to get ($\a\equiv\a(s_0)$)
\begin{eqnarray}
\label{Iint}
I(s_0,p)&=&(-1)^{p+1} \p \a + \sum_{n\geq 3}\frac{(-1)^{p+1} \p^n \sin(n \p/2)}{n!}\int_0^\infty d\o\, e^{-\o/\a} (\o-p)^{n-1} \\
&=&(-1)^{p+1} \p\a + e^{-p/\a}\, \sum_{n\geq 3}\frac{(-1)^{p+1} \p^n \sin(n \p/2)}{n!}\, \a^n \, \G(n,-p/\a)\ .\nonumber
\end{eqnarray}
Given that $\G(n,-p/\a)\approx (n-1)!$ as $n\to \infty$, the radius of covergence is that of the series
\begin{equation}
\sum_{n\geq 3} \frac{\p^n \sin(n \p/2)}{n}\, \a^n = \sum_{k\geq 1} (-1)^k  \frac{(\p\a)^{2k+1}}{2k+1} = \p\a \sum_{k\geq 1} (-1)^k  \frac{(\p\a)^{2k}}{2k+1}  \ ,
\end{equation}
which results in a series in $\a^2$ which is convergent for $\a^2<1/\pi^2$, or $\a<1/\p$. Interestingly the radius of convergence does not depend on $p$ or $m$.

\vskip0.8cm
\begin{boldmath}
\section{\label{OtherAS} Another derivation of the AS}
\end{boldmath}

An alternative derivation of the AS results in Eq.~(\ref{ASlargebetaResult}) is the following. For simplicity, we will consider only the simple pole case, $\g=1$. Starting from the CIPT series for the moment $A_m(s_0)
=\sum_{n=1}^\infty n!\,A_{m,n}(s_0)$ one has the expression Eq.~(\ref{FOPTmaster}) written in terms of the $I_T$ integral in (\ref{masterT}), instead of the Borel integral form $I$ in Eq.~(\ref{master}). Making the change of variables $\o'=\o (1\pm i\p\a)$ we obtain ($\e>0$)\footnote{We could follow the standard rule of the Principal Value prescription and average over $\pm i\e$ but this is not necessary.}
\begin{eqnarray}
\label{AmnCIPTseries3}
n!\,A_{m,n}(s_0)&=&\frac{1}{2 i\p p}\ \frac{1}{(1+i\p\a)^{n+1}}\int_{\Gamma}
d\o'\,\left(\frac{\o'}{p}\right)^n\,\frac{e^{-\o' /\a}}{\o'/(1+i\p\a)-m-i\e}\\
&-&\frac{1}{2 i\p p}\ \frac{1}{(1-i\p\a)^{n+1}}\int_{\Gamma'}
d\o'\,\left(\frac{\o'}{p}\right)^n\,\frac{e^{-\o' /\a}}{\o'/(1-i\p\a)-m-i\e}\ ,\nonumber
\end{eqnarray}
where, as a consequence of the change of variables,  the integral no longer is along the positive real axis but is along a straight line $\G'$ which  runs from zero to infinity making
an angle $-\arctan(\p\a)$ with the positive real axis, or a contour $\G$ which runs from zero to infinity making
an angle $\arctan(\p\a)$ with the positive real axis.

These two contours $\Gamma^{(')}$ may now be rotated to the positive real axis, picking up a contribution from the pole at $(m+i\e)(1\pm i\p\a)$, and yielding:
\begin{eqnarray}
\label{AmnCIPTseries4}
n!\,A_{m,n}(s_0)&=&\frac{1}{2 i\p p}\ \frac{1}{(1+i\p\a)^{n+1}}\int_{0}^\infty
d\o'\,\left(\frac{\o'}{p}\right)^n\,\frac{e^{-\o' /\a}}{\o'/(1+i\p\a)-m-i\e}\nonumber\\
&-&\frac{1}{2 i\p p}\ \frac{1}{(1-i\p\a)^{n+1}}\int_{0}^\infty
d\o'\,\left(\frac{\o'}{p}\right)^n\,\frac{e^{-\o' /\a}}{\o'/(1-i\p\a)-m-i\e}\nonumber\\
&&\hspace{1cm}-(-1)^m\,e^{-m/\a}\frac{1}{p}\left(\frac{m}{p}\right)^n\ .
\end{eqnarray}
One can now define the Borel sum by summing over $n$ to obtain
\begin{eqnarray}
\label{AmnCIPTSeries5}
A^{\rm CIPT}_m(s_0)&=&\sum_{n=1}^\infty n!\,A_{m,n}(s_0)\\
&=&\frac{1}{2i\p}\,\int_0^\infty
\frac{d\o'}{1+i\p\a}\,\frac{1}{p-\frac{\o'}{1+i\p\a}}\,\frac{e^{-\o' /\a}}{\frac{\o'}{1+i\p\a}-m-i\e}\nonumber\\
&-& \frac{1}{2i\p}\,\int_0^\infty
\frac{d\o'}{1-i\p\a}\,\frac{1}{p-\frac{\o'}{1-i\p\a}}\,\frac{e^{-\o' /\a}}{\frac{\o'}{1-i\p\a}-m-i\e}\nonumber \\
&&-\frac{(-1)^m}{p-m}\,e^{-m/\a}\ .\nonumber
\end{eqnarray}

Undoing the previous change of variables $\o=\o'/(1\pm i \p\a)$ results in
\begin{eqnarray}
\label{AmnCIPTSeries6}
A^{\rm CIPT}_m(s_0)
&=&\frac{1}{2\p i}\int_{\G'}
d\o\,\frac{ e^{-i\p\o}}{(p+i\e-\o)(\o-m-i\e)}\,e^{-\o/\a}\\
&&-\frac{1}{2\p i}\int_{\G}
d\o\,\frac{ e^{i\p\o}}{(p+i\e-\o)(\o-m-i\e)}\,e^{-\o/\a}\nonumber\\
&&-\frac{(-1)^m}{p-m}\,e^{-m/\a}\ ,\nonumber
\end{eqnarray}
where we have also regulated the pole at $p+i\e$, as discussed in the main text.

It is intuitively clear that this back and forth of changing of variables  is a way to detect the difference between a function with a pole and the same function as a power series, through the help of the residue theorem.

We can now look at the FOPT-CIPT difference:
\begin{eqnarray}
\label{FOPTmCIPT}
A^{\rm FOPT}_m(s_0)-A^{\rm CIPT}_m(s_0)&=&
\frac{1}{2\p i}\int_{C'}
d\o\,\frac{e^{-i\p\o}}{(p+i\e-\o)(\o-m-i\e)}\,e^{-\o/\a}\\
&&-\frac{1}{2\p i}\int_{C}
d\o\,\frac{e^{i\p\o}}{(p+i\e-\o)(\o-m-i\e)}\,e^{-\o/\a}\nonumber\\
&&+\frac{(-1)^m}{p-m}\,e^{-m/\a}\ ,\nonumber
\end{eqnarray}
where $C$ is the contour from 0 to $\infty$ and back along $\G$
(which is clockwise),
$C'$ is the contour from 0 to $\infty$ and back along $\G'$
(which is counter-clockwise),
and we regulated the poles at $\o=p$ and $\o=m$. We
obtain
\begin{equation}
\label{FOPTmCIPT1}
A^{\rm FOPT}_m(s_0)-A^{\rm CIPT}_m(s_0)=(-1)^p
\frac{1}{p-m}\,e^{-p/\a}\ .
\end{equation}
One can verify that choosing $\e<0$ gives the
same result.   The contribution from the pole at $\o=m$ is cancelled by the
term on the third line of Eq.~(\ref{FOPTmCIPT}).   This is the result in the first Eq.~(\ref{resultAS}).

%%%%%%%%%%%%%%%%%%%%%%%%%%%

\end{document}